\documentclass{article}
\usepackage{spconf,amsmath,graphicx}

\usepackage{multirow}
\usepackage{amssymb}

\usepackage{graphicx}
\usepackage{subcaption}
\usepackage{cite}
\usepackage{colortbl} 
\usepackage{color}

\title{Neural Architecture Search For Speech Emotion Recognition}
\name{Xixin Wu$^1$\sthanks{
This research is supported by the CUHK Stanley Ho Big Data Decision Analytics Research Centre, the Centre for Perceptual and Interactive Intelligence, and National Natural Science Foundation of China (62076144).}, Shoukang Hu$^1$, Zhiyong Wu$^{1,2}$, Xunying Liu$^1$, Helen Meng$^1$}
\address{$^1$ The Chinese University of Hong Kong, Hong Kong SAR, China\\
$^2$ Shenzhen International Graduate School, Tsinghua University, Shenzhen, China}
\begin{document}
%
\maketitle
\begin{abstract}
Deep neural networks have brought significant advancements to speech emotion recognition (SER). 
However, the architecture design in SER is mainly based on expert knowledge and empirical (trial-and-error) evaluations, which is time-consuming and resource intensive.
In this paper, we propose to apply neural architecture search (NAS) techniques to automatically configure the SER models. 
To accelerate the candidate architecture optimization, we propose a uniform path dropout strategy to encourage all candidate architecture operations to be equally optimized. 
Experimental results of two different neural structures on IEMOCAP show that NAS can improve SER performance (54.89\% to 56.28\%) while maintaining model parameter sizes. The proposed dropout strategy also shows superiority over the previous approaches.
\end{abstract}
\section{Introduction}
\label{sec:intro}
Speech emotion recognition (SER) is an important contributor towards graceful human-machine interaction, as machines can generate appropriate responses according to human emotions in the contexts of interaction.  With the applications of deep neural networks, SER research has made great progresses in the last decade~\cite{han2014speech,lee2015high,li2018attention,wu2019speech,fujioka2020meta,wang2020speech}.  The architecture designing of neural models in SER is generally reliant on expert knowledge and empirical (trial-and-error) evaluations. As explicit training and evaluation of model designs are often time-consuming and resource intensive, it becomes worthwhile to explore neural architecture search (NAS) techniques to automate the neural architecture design process.
Hence, the field of NAS has attracted much research attention recently \cite{zoph2016neural, pham2018efficient, floreano2008neuroevolution, real2017large, real2019regularized, liu2018darts}.
Generally, an NAS algorithm is designed to search architectures by combining operations from a predefined space based on certain evaluation metrics.  
The approaches based on reinforcement learning (RL) \cite{zoph2016neural, pham2018efficient}, evolution \cite{angeline1994evolutionary, floreano2008neuroevolution, real2017large, real2019regularized} or Bayesian optimization \cite{shahriari2015taking} demonstrate outstanding performance.  However, these methods demand huge computational resources for system training and evaluation (e.g., 1800 GPU days \cite{zoph2016neural}).
Instead of searching over a discrete architecture space, the differentiable architecture search (DARTS) approach allows the space to be continuous, i.e., the categorical network operation choice is replaced with a softmax over all possible operations \cite{liu2018darts}.  
The architecture search task is then transformed to the learning of the softmax function outputs with gradient descent optimization, which significantly reduces the training costs (e.g., 4 GPU days \cite{liu2018darts}).  The over-parameterized network, referred to as \textit{supernet}, is composed of the combinations of operations with the architecture weights, i.e., the softmax outputs. 
As the candidate operation parameters and the architecture weights are optimized on the same supernet, the sub-optimal architectures may be prematurely learned at an early stage of the training.  
To encourage all candidate architectures to be optimized simultaneously, various techniques are applied, e.g., operation dropout \cite{bender2018understanding}, uniform path sampling \cite{guo2020single,hu2021neural}.  
However, the training using operation dropout is often not stable, since certain nodes in the supernet may drop all operations. The training efficiency of uniform path sampling is low because at each training step only one chosen operation 
is optimized.  

The NAS techniques have also been successfully applied to speech synthesis \cite{luo2021lightspeech} and speech recognition \cite{moriya2018evolution, kim2020evolved, chen2020darts, he2021learned, zheng2021efficient, hu2021neural, liu2021improved, shi2021darts}. It has been shown that the performance can be improved and the model parameter sizes can be reduced at the same time. However, limited NAS research has been conducted in the SER area.  
In this paper, we investigate differentiable NAS for SER based on the two effective systems with the attention mechanism \cite{li2018attention} and capsule networks \cite{wu2019speech}, respectively.   
To further improve the efficient training of the candidate operations, we propose a simple yet effective strategy, \textit{uniform path dropout}.  Different from the sampling strategy that selects only one single path each time, our strategy drops a group of paths and selects the rest, so that the optimization can be accelerated, as multiple paths are optimized each time.  Additionally, dropping paths also destroys the supernet's reliance on sub-optimal operations and forces the supernet to optimize other operations.
Experimental results show that the proposed approach can improve emotion recognition performance by maintaining a moderate model parameter size. As far as we know, this is among the first efforts to investigate the effectiveness of NAS for the SER task.


\section{Differentiable Neural Architecture Search}
\label{sec:dnas}
\begin{figure}
    \centering
    \includegraphics[width=0.5\textwidth]{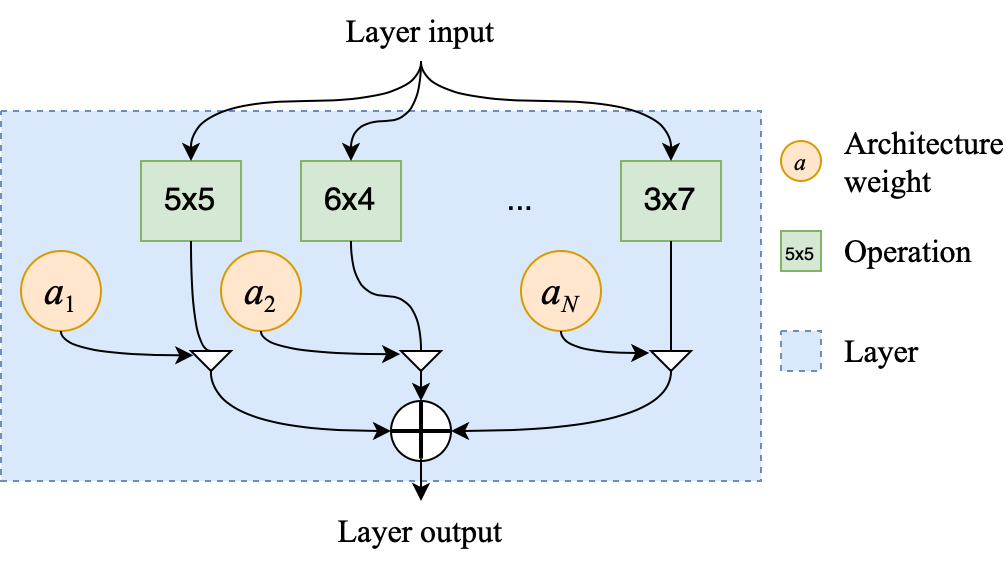}
    \caption{A layer in the supernet is composed of candidate operations and the associated architecture weights.  The operations, such as convolutional layers with different kernel sizes, are combined based on the architecture weights.}
    \label{fig:node}
\end{figure}

The neural architecture search (NAS) techniques aim at automatically selecting neural architectures given a predefined search space, search algorithm and evaluation metrics \cite{zoph2018learning,liu2018darts}. In this paper, we focus on the branch of differentiable architecture search (DARTS), which relaxes the search space to be continuous so that gradient descent optimization can be applied to the over-parameterized supernet.  The supernet is built by connecting the layers that combine all the candidate operations, e.g., convolutional layers with different kernel sizes, in the search space, as shown in Fig~\ref{fig:node}.  The operations in the layer are weighted combined as:
\begin{eqnarray}
\mathbf{h}^l=\sum_{i=1}^{N^l}a^l_i\phi_i(\mathbf{h}^{l-1};\mathbf{W}^l_i),
\end{eqnarray}
where $\mathbf{a}^l=\{a^l_1, a^l_2,...,a^l_{N^l}\}$ is the architecture weights for the $N^l$ candidate operations in the $l$-th layer. $\mathbf{h}^{l-1}$ and $\mathbf{h}^l$ are the layer input and output, respectively. $\phi_i$ is the $i$-th operation with the model parameters $\mathbf{W}_i^l$.
Possible options for modeling the architecture weights in the supernet are using the softmax function or conducting proximal iterations \cite{yao2020efficient}.

\subsection{Joint Optimization}
The NAS can be performed by jointly optimizing the architecture weights $\mathcal{A}=\{\mathbf{a}^l\}_{l=1}^M$ and model parameters $\mathcal{W}=\{\mathbf{W}^l\}_{l=1}^M$ based on the training data:
\begin{eqnarray}
\mathcal{A}^*,\mathcal{W}^*=\arg\min_{\mathcal{A},\mathcal{W}} \mathcal{L}_{\tt train}(\mathbf{N}(\mathcal{A},\mathcal{W})),
\end{eqnarray}
where $M$ is the number of layers, $\mathcal{L}_{\tt train}(\mathbf{N}(\mathcal{A},\mathcal{W}))$ is training loss of the supernet $\mathbf{N}(\mathcal{A},\mathcal{W})$.  The final architecture is selected by connecting the operations corresponding to maximum architecture weights, which exhibits a connection to the network pruning techniques \cite{han2015learning}.  Jointly training the architecture weights and the model parameters saves time.  However, the sub-optimal operations (e.g., simpler operations) may dominate the weights at an early stage, such that the optimal operations (e.g., operations with larger parameter sizes) may be ignored \cite{hu2021neural}.

\subsection{Bi-level Optimization}
Alternatively, a bi-level optimization can be used for NAS:
\begin{eqnarray}
&\mathcal{A}^* = \arg\min_{\mathcal{A}}\mathcal{L}_{\tt val}(\mathbf{N}(\mathcal{A},\mathcal{W}^*)), \label{eq:a_optimization}\\
&\text{s.t.}~~\mathcal{W}^*=\arg\min_{\mathcal{W}}\mathcal{L}_{\tt train}(\mathbf{N}(\mathcal{A},\mathcal{W})), \label{eq:w_optimization}
\end{eqnarray}
where $\mathcal{L}_{\tt val}$ is the validation loss. This decouples the training of the architecture weights and the model parameters.
To solve this problem, \cite{liu2018darts} approximates the optimal model parameters in Eq.~(\ref{eq:a_optimization}) via an one-step forward update.  However, the learning rate for the one-step update needs to be carefully chosen.
Alternatively, the uniform path sampling strategy is adopted for Eq.~(\ref{eq:w_optimization}) by \cite{guo2020single,hu2021neural}.  The idea is to randomly select one path in the supernet (i.e. one candidate architecture) and then optimize the weights of the operations along the path. In this way, the weights of all candidate architectures are optimized by optimizing the supernet. However, since only the parameters of one architecture are optimized at each step, the training converges slowly.  
\begin{figure}
    \centering
    \includegraphics[width=0.35\textwidth]{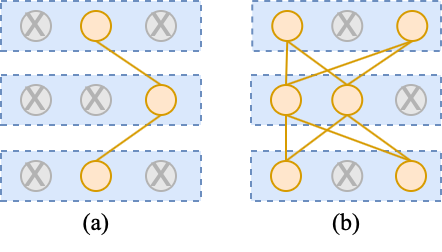}
    \caption{Comparison of (a) uniform path sampling and (b) uniform path dropout.  Only one path is selected via path sampling, while a group of paths is selected via path dropout.}
    \label{fig:path_drop}
\end{figure}

To increase training efficiency, we propose to adopt a uniform path dropout strategy to randomly drop a group of paths and select the remaining paths, instead of selecting only one single path, as shown in Fig.~\ref{fig:path_drop}.  More specifically, by randomly masking a constant fraction (e.g., 1 out of 6) of operations in one layer, all the paths going through the masked operations are dropped, and the other paths are selected.  The layer output is scaled up according to the mask fraction for training stability \cite{srivastava2014dropout}.  
Selecting multiple paths aims at accelerating training. Dropping paths is necessary for destroying the supernet's reliance on certain sub-optimal operations and encouraging simultaneous optimization of the operation parameters (more discussion in Section~\ref{sec:exp_result}).
Different from the strategy by \cite{han2015learning}, where each operation is independently dropped, our dropout strategy is performed on the layer level and for each step a constant number of operations are dropped, so the risk of all operations are dropped can be avoided \cite{guo2020single}.  In our experiments, we found that our strategy is more stable during training because of constant number of dropout.
\section{System Architecture}
\begin{table}[t]
 \begin{center}
  \caption{Configuration of the CNN component, attention layer, dense layer and the candidate operations for NAS.  C, D, K, and W stand for channel number, dimension, kernel size and pooling window, respectively.} \label{tab:cnn}
  \begin{tabular}{c|c|c}
   \hline
   Layer & Structure & Operations \\
   \hline
   \hline
   Conv2d\_1 & C=8, K=2$\times$8 & 2$\times$8,2$\times$7,2$\times$6,\\
   Conv2d\_2 & C=8, K=8$\times$2 & 1$\times$9,1$\times$10,3$\times$5\\
   \hline
   Concat & Conv2d\_1 + Conv2d\_2 & \multirow{2}{*}{-}\\
   Max-pooling & W=2$\times$1 &  \\
   \hline
   \multirow{2}{*}{Conv2d\_3} & \multirow{2}{*}{C=16, K=5$\times$5} & 5$\times$5,5$\times$4,4$\times$5,\\
   & &  4$\times$4,4$\times$6,6$\times$4 \\
   \hline
   Max-pooling & W=2$\times$2 & - \\
   \hline
   \multirow{2}{*}{Conv2d\_4} & \multirow{2}{*}{C=16, K=5$\times$5} & 5$\times$5,5$\times$4,4$\times$5,  \\
   & & 4$\times$4,4$\times$6,6$\times$4 \\
   \hline
   Max-pooling & W=2$\times$2 & \multirow{2}{*}{-} \\
   Max-pooling & W=4$\times$1 &   \\
   \hline
   \hline
   Attention & \multirow{2}{*}{C=64} & \multirow{2}{*}{32,48,64,80}\\
   (CNN\_RNN\_att) & & \\
   \hline
   \hline
   Dense & \multirow{2}{*}{D=64} & \multirow{2}{*}{32,48,64,80}\\
   (CNN\_SeqCap) & & \\
   \hline
  \end{tabular}
 \end{center}
\end{table}
One representative among the various successful network structures for SER is convolutional neural networks (CNNs), which have demonstrated the effectiveness via learning neural hidden representations directly from spectrograms or waveforms \cite{tzirakis2018end,wu2019speech}.  However, the convolutional layer configurations are designed mainly based on expert experience and evaluations.  In this section, we describe the application of the NAS methods introduced in Section~\ref{sec:dnas} to optimize the two systems of CNN\_GRU\_att \cite{li2018attention} and CNN\_SeqCap \cite{wu2019speech}. 

The CNN\_GRU\_att and the CNN\_SeqCap both have the CNN module consisting of multiple convolutional and pooling layers to extract neural representations from spectrograms for the subsequent layers.  Table \ref{tab:cnn} presents the configuration of the CNN module.
Two separated convolutional layers with kernel of 2$\times$8 and 8$\times$2 are adopted.  The outputs of these two separated layers are concatenated and passed through another two convolutional layers and three max-pooling layers. For the convolutional layers, we define three sets of candidate operations for the search space.  The operations are convolutional layers with different kernel sizes, as shown in Table \ref{tab:cnn}.  We intend to include smaller kernel sizes in the sets as one of our goal is to reduce the model parameter sizes.

Upon the CNN module, the CNN\_GRU\_att system has a bi-directional gated recurrent unit (GRU) layer with 64 cells per direction.  The final state of foward GRU and the first state of backward GRU are concatenated and fed to an attention layer that is composed of class-agnostic bottom-up and class-specific top-down attention maps \cite{girdhar2017attentional}.  The channel number for the attention layer adopted in \cite{wu2019speech} is 64.  We define the set of \{32, 48, 64, 80\} as candidate channel numbers of the attention layer for search. 

For the CNN\_SeqCap system, a sequential capsule layer is applied to the CNN module outputs \cite{wu2019speech}. 
These capsules obtained from 8 convolutional layers are routed to window-level capsule layer with 8 capsules of size 8 in each window.  An utterance-level routing is conducted upon the window output vectors to produce 4 utterance-level capsules of size 16. The utterance-level capsules are then fed to two dense layers and softmax function.  We define the set of \{32, 48, 64, 80\} for the candidate dimensions of the first dense layer.   The window used to slice the CNN module outputs is set to size of 40 input steps with shift of 20 steps.  The routing iteration number is set to 3, and the number of masked operations is set to 1.

\section{Experiments And Analysis}
\subsection{Emotion Recognition Corpus}
We conducted experiments on the benchmark corpus IEMOCAP \cite{busso2008iemocap}, which consists of five sessions with two speakers in each session.  The five-fold cross validation is adopted as \cite{satt2017efficient,wu2019speech}: 8 speakers from four sessions in the corpus are used as training data. One speaker from the remaining session is used as validation data, and the other one as test data.  For NAS, the training data is used to train the model parameters and the validation data is used for optimizing the architecture weights.
We evaluate on four emotions of the improvised data in the corpus, i.e., \textit{Neutral}, \textit{Angry}, \textit{Happy} and \textit{Sad}, following previous work \cite{ma2018emotion,satt2017efficient}.  
Spectrograms are extracted from the speech signal and split into 2-second segments, sharing the same utterance-level label.
The training is conducted based on the 2-second segments, and the whole original spectrogram is used for evaluation. The spectrograms are extracted with 40-ms Hanning window, 10-ms shift and discrete Fourier transform (DFT) of length 16k, and normalized to have zero mean and unit variance.

\subsection{Training Configuration}
The Xavier initializer is used for the initialization of both the CNN components and the capsule layers.  For the joint optimization mode, the supernet is trained for 60 epochs using the Adam algorithm ($\beta_1=0.9$, $\beta_2=0.999$ and $\epsilon=$1$e$-8) \cite{kingma2015adam} with a dynamic learning rate scheme \cite{wu2019speech}. The learning rate is set to 0.001 in the first 3 epochs and reduced to 0.0005, 0.0002 and 0.0001 gradually, when the average training loss is reduced by a factor of 10. The batch size is set to 16. 
For the bi-level optimization, the supernet is trained in the same way as the joint optimization, and the architecture weights are trained on the validation set using a constant learning rate of 0.001.
The selected architecture is then trained again for 20 epochs and then optimized on the validation set with respect to the weighted accuracy.  

\subsection{Experimental Results}
\label{sec:exp_result}
\begin{table}
    \centering
    \begin{tabular}{c|c|c c|c}
        \hline
        System & NAS & WA(\%) & UA(\%) & \#param(k)  \\
        \hline
        \multirow{5}{*}{CNN\_RNN\_att} & $\times$ & 68.20 & 54.89 & 833  \\
        & Random & 68.42 & 54.81 & 833  \\
        \cline{2-5}
        & Joint & 68.61 & 55.59 & 833 \\
        & Sampling & \textbf{69.10} & 54.23 & 835 \\
        & Dropout & 68.87 & \textbf{56.28}& \textbf{832} \\
        \hline
        \hline
        \multirow{5}{*}{CNN\_SeqCap} & $\times$ & 69.86 & 56.71& 704\\
        & Random & 68.77& 54.75 & 700\\
        \cline{2-5}
        & Joint & 68.80 & 54.44 & 704\\
        & Sampling & 70.18 & 56.72 & 704\\
        & Dropout & \textbf{70.54} & \textbf{56.94} & \textbf{700}\\
        \hline
    \end{tabular}
    \caption{Performance of the proposed and baseline systems.}
    \label{tab:exp}
\end{table}
We use two common evaluation metrics to evaluate system performance, i.e., weighted accuracy (WA) and unweighted accuracy (UA).  
WA is the accuracy of all samples in the test data, and UA is the average of class accuracies in the test set.  
The performance comparison of architectures found by NAS and the baseline systems are shown in Table~\ref{tab:exp}.  The random search, which selects the best performed architecture from 5 randomly sampled architectures, turns out to be a considerably strong baseline that achieves comparable performance with the system found by joint optimization.  This coincides with the observations in previous literature \cite{liu2018darts,hu2021neural}.  We need to also note that the baseline architectures are already optimized by the authors, which also provides a sufficiently good starting point for the random search.  Table~\ref{tab:exp} also shows that both the uniform path sampling and dropout strategies are effective in boosting the performance to outperform the original models (without NAS) and the random search baselines.  The proposed dropout strategy is better than the sampling strategy in terms of recognition performance.  While the joint optimization is inferior to the two bi-level optimization strategies, it can still achieve better or comparable performance with the random search method.

\begin{figure}[h]
    \centering
    \includegraphics[width=0.4\textwidth]{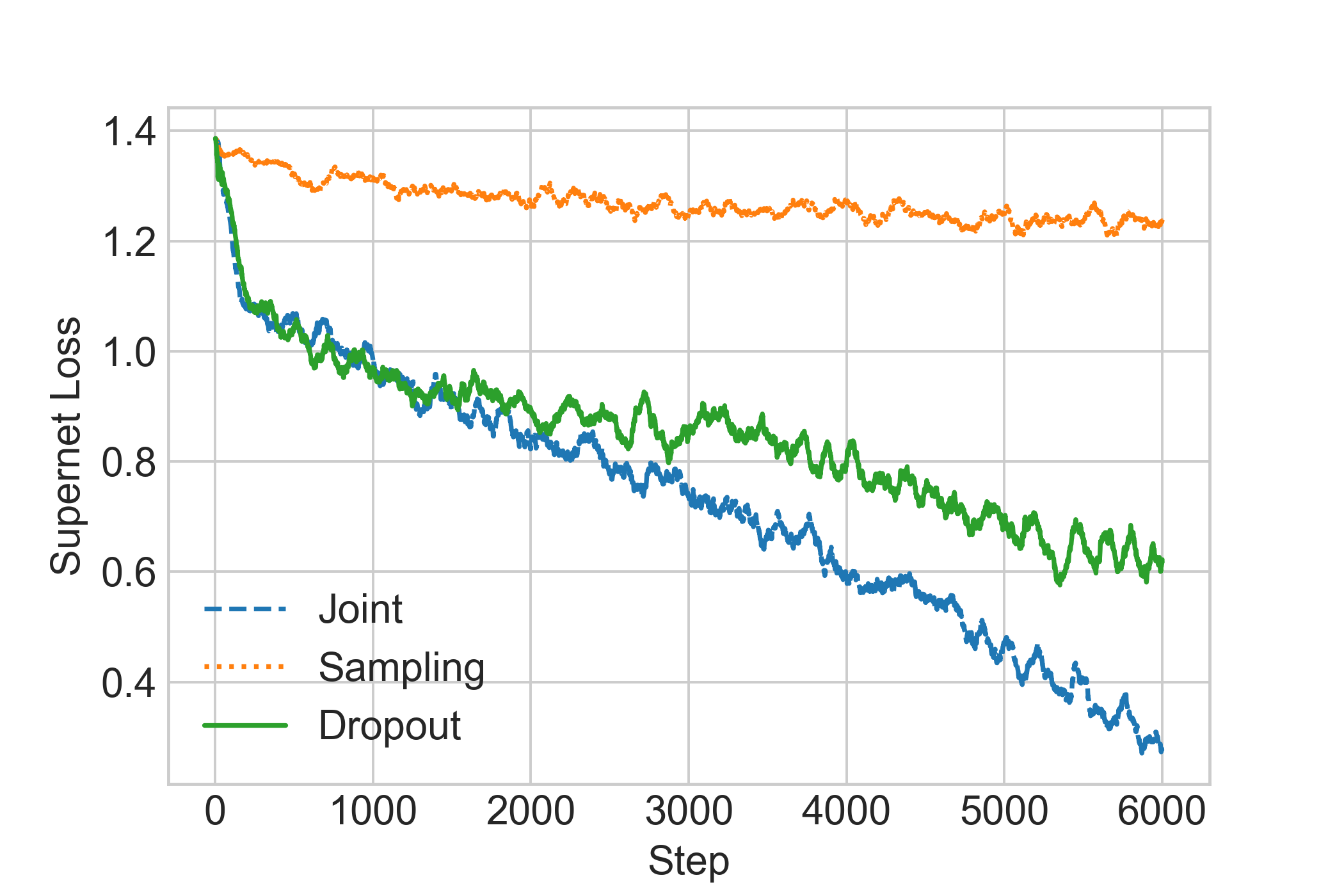}
    \centerline{(a) Supernet loss}
    \includegraphics[width=0.4\textwidth]{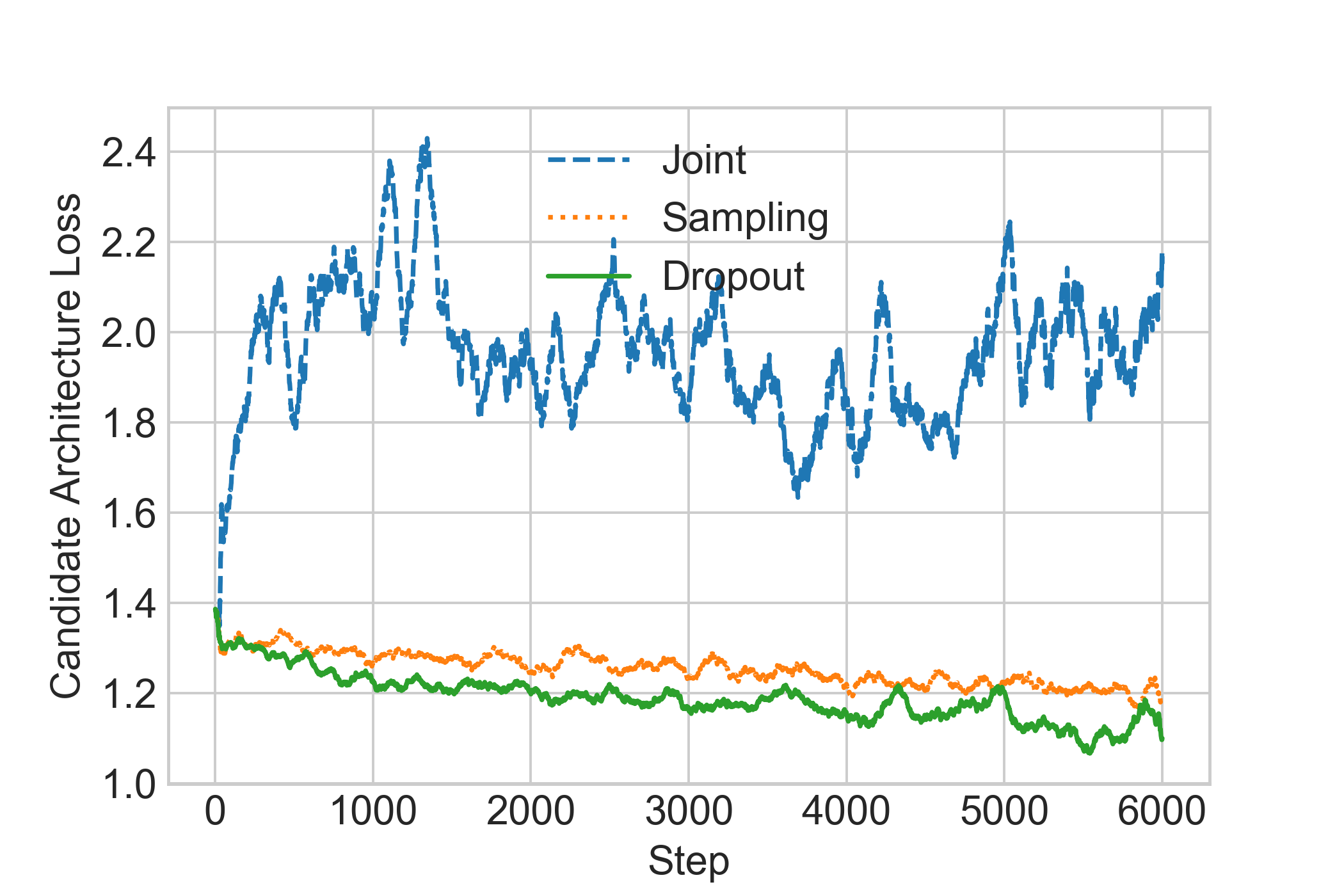}
    \centerline{(b) Candidate architecture loss}
    \caption{Training loss curves of (a) the supernet and (b) a randomly selected candidate architecture of CNN\_SeqCap. 
    }
    \label{fig:loss}
\end{figure}
To further analyze the behaviours of the sampling and dropout strategy, we plot the training loss curves of the supernet in Fig.~\ref{fig:loss} (a) and a candidate architecture randomly selected from the supernet in Fig.~\ref{fig:loss} (b).  It is verified that the proposed path dropout strategy can accelerate the training via selecting multiple paths at each training step, as shown in Fig.~\ref{fig:loss} (a) that the dropout loss curve is closer to the loss curve of joint optimization.  While the sampling strategy slows down the decrease of the supernet loss, the candidate architecture loss curve is much lower than the joint optimization curve as shown in Fig.~\ref{fig:loss} (b).  This implies that the sampling strategy can simultaneously optimize the parameters of various candidate operations.  The dropout strategy produces an even lower loss curve than the sampling one, which demonstrates the strategy's superiority and provides insights for the performance gains achieved, as shown in Table~\ref{tab:exp}.

\section{Conclusions}
In this paper, we investigate neural architecture search for speech emotion recognition to boost the recognition performance with reduced model parameter sizes.  Experimental results suggest the effectiveness of the baseline and proposed search strategies.  It is demonstrated that the uniform path dropout strategy can encourage all architecture operations to be equally optimized by randomly dropping paths, and increase the training efficiency by selecting multiple paths at each step. In the future, we will investigate more diverse search spaces and transferability of found architectures.


\ninept
\bibliographystyle{IEEEbib}
\bibliography{refs}

\begin{thebibliography}{10}

\bibitem{han2014speech}
K.~Han, D.~Yu, and I.~Tashev,
\newblock ``Speech emotion recognition using deep neural network and extreme
  learning machine,''
\newblock in {\em Annual Conference of ISCA}, 2014.

\bibitem{lee2015high}
J.~Lee and I.~Tashev,
\newblock ``High-level feature representation using recurrent neural network
  for speech emotion recognition,''
\newblock in {\em Annual Conference of the ISCA}, 2015.

\bibitem{li2018attention}
P.~Li, Y.~Song, I.~McLoughlin, W.~Guo, and L.~Dai,
\newblock ``An attention pooling based representation learning method for
  speech emotion recognition,''
\newblock in {\em Proc. of INTERSPEECH}, 2018, pp. 3087--3091.

\bibitem{wu2019speech}
X.~Wu, S.~Liu, Y.~Cao, X.~Li, J.~Yu, D.~Dai, X.~Ma, S.~Hu, Z.~Wu, X.~Liu, and
  H.~Meng,
\newblock ``Speech emotion recognition using capsule networks,''
\newblock in {\em Proc. of ICASSP}, 2019, pp. 6695--6699.

\bibitem{fujioka2020meta}
T.~Fujioka, T.~Homma, and K.~Nagamatsu,
\newblock ``Meta-learning for speech emotion recognition considering ambiguity
  of emotion labels.,''
\newblock in {\em Proc. of INTERSPEECH}, 2020, pp. 2332--2336.

\bibitem{wang2020speech}
J.~Wang, M.~Xue, R.~Culhane, E.~Diao, J.~Ding, and V.~Tarokh,
\newblock ``Speech emotion recognition with dual-sequence {LSTM}
  architecture,''
\newblock in {\em ICASSP}, 2020, pp. 6474--6478.

\bibitem{zoph2016neural}
B.~Zoph and Q.~Le,
\newblock ``Neural architecture search with reinforcement learning,''
\newblock {\em ICLR}, 2017.

\bibitem{pham2018efficient}
H.~Pham, M.~Guan, B.~Zoph, Q.~Le, and J.~Dean,
\newblock ``Efficient neural architecture search via parameter sharing,''
\newblock {\em ICML}, pp. 4095--4104, 2018.

\bibitem{floreano2008neuroevolution}
D.~Floreano, P.~D{\"u}rr, and C.~Mattiussi,
\newblock ``Neuroevolution: from architectures to learning,''
\newblock {\em Evolutionary intelligence}, vol. 1, no. 1, pp. 47--62, 2008.

\bibitem{real2017large}
E.~Real, S.~Moore, A.~Selle, and et~al.,
\newblock ``Large-scale evolution of image classifiers,''
\newblock in {\em ICML}, 2017, pp. 2902--2911.

\bibitem{real2019regularized}
E.~Real, A.~Aggarwal, Y.~Huang, and Q.~V Le,
\newblock ``Regularized evolution for image classifier architecture search,''
\newblock in {\em Proc. of AAAI}, 2019, vol.~33, pp. 4780--4789.

\bibitem{liu2018darts}
H.~Liu, K.~Simonyan, and Y.~Yang,
\newblock ``Darts: Differentiable architecture search,''
\newblock in {\em Proc. of ICLR}, 2018.

\bibitem{angeline1994evolutionary}
P.~Angeline, G.~Saunders, and J.~Pollack,
\newblock ``An evolutionary algorithm that constructs recurrent neural
  networks,''
\newblock {\em IEEE transactions on Neural Networks}, vol. 5, no. 1, pp.
  54--65, 1994.

\bibitem{shahriari2015taking}
B.~Shahriari, K.~Swersky, Z.~Wang, R.~Adams, and De~F.,
\newblock ``Taking the human out of the loop: A review of bayesian
  optimization,''
\newblock {\em Proceedings of the IEEE}, vol. 104, no. 1, pp. 148--175, 2015.

\bibitem{bender2018understanding}
G.~Bender, P.~Kindermans, B.~Zoph, V.~Vasudevan, and Q.~Le,
\newblock ``Understanding and simplifying one-shot architecture search,''
\newblock in {\em ICML}. PMLR, 2018, pp. 550--559.

\bibitem{guo2020single}
Z.~Guo, X.~Zhang, H.~Mu, W.~Heng, Z.~Liu, Y.~Wei, and J.~Sun,
\newblock ``Single path one-shot neural architecture search with uniform
  sampling,''
\newblock in {\em Proc. of ECCV}, 2020, pp. 544--560.

\bibitem{hu2021neural}
S.~Hu, X.~Xie, S.~Liu, M.~Cui, M.~Geng, X.~Liu, and H.~Meng,
\newblock ``Neural architecture search for lf-mmi trained time delay neural
  networks,''
\newblock in {\em Proc. of ICASSP}. IEEE, 2021, pp. 6758--6762.

\bibitem{luo2021lightspeech}
R.~Luo, X.~Tan, R.~Wang, T.~Qin, J.~Li, S.~Zhao, E.~Chen, and T.~Liu,
\newblock ``Lightspeech: Lightweight and fast text to speech with neural
  architecture search,''
\newblock in {\em Proc. of ICASSP}. IEEE, 2021, pp. 5699--5703.

\bibitem{moriya2018evolution}
T.~Moriya, T.~Tanaka, T.~Shinozaki, S.~Watanabe, and K.~Duh,
\newblock ``Evolution-strategy-based automation of system development for
  high-performance speech recognition,''
\newblock {\em IEEE/ACM Transactions on Audio, Speech, and Language
  Processing}, vol. 27, no. 1, pp. 77--88, 2018.

\bibitem{kim2020evolved}
K.~Jihwan, W.~Jisung, K.~Sangki, and et~al.,
\newblock ``Evolved speech transformer: Applying neural architecture search to
  end-to-end automatic speech transformer,''
\newblock {\em INTERSPEECH}, pp. 1788--1792, 2020.

\bibitem{chen2020darts}
Y.~Chen, J.~Hsu, C.~Lee, and H.~Lee,
\newblock ``Darts-asr: Differentiable architecture search for multilingual
  speech recognition and adaptation,''
\newblock {\em INTERSPEECH}, pp. 1803--1807, 2020.

\bibitem{he2021learned}
L.~He, D.~Su, and D.~Yu,
\newblock ``Learned transferable architectures can surpass hand-designed
  architectures for large scale speech recognition,''
\newblock in {\em ICASSP}, 2021, pp. 6788--6792.

\bibitem{zheng2021efficient}
H.~Zheng and et~al.,
\newblock ``Efficient neural architecture search for end-to-end speech
  recognition via straight-through gradients,''
\newblock in {\em IEEE Spoken Language Technology Workshop (SLT)}. IEEE, 2021,
  pp. 60--67.

\bibitem{liu2021improved}
Y.~Liu, T.~Li, P.~Zhang, and Y.~Yan,
\newblock ``Improved conformer-based end-to-end speech recognition using neural
  architecture search,''
\newblock {\em arXiv preprint arXiv:2104.05390}, 2021.

\bibitem{shi2021darts}
X.~Shi, P.~Zhou, W.~Chen, and L.~Xie,
\newblock ``Darts-conformer: Towards efficient gradient-based neural
  architecture search for end-to-end asr,''
\newblock {\em arXiv preprint arXiv:2104.02868}, 2021.

\bibitem{zoph2018learning}
B.~Zoph, V.~Vasudevan, J.~Shlens, and Q.~V Le,
\newblock ``Learning transferable architectures for scalable image
  recognition,''
\newblock in {\em Proc. of CVPR}, 2018, pp. 8697--8710.

\bibitem{yao2020efficient}
Q.~Yao, J.~Xu, W.~Tu, and Z.~Zhu,
\newblock ``Efficient neural architecture search via proximal iterations,''
\newblock in {\em Proc. of AAAI}, 2020, vol.~34, pp. 6664--6671.

\bibitem{han2015learning}
S.~Han, J.~Pool, J.~Tran, and W.~Dally,
\newblock ``Learning both weights and connections for efficient neural
  network,''
\newblock {\em Proc. of NeurIPS}, vol. 28, 2015.

\bibitem{srivastava2014dropout}
N.~Srivastava, G.~Hinton, A.~Krizhevsky, I.~Sutskever, and R.~Salakhutdinov,
\newblock ``Dropout: a simple way to prevent neural networks from
  overfitting,''
\newblock {\em The Journal of Machine Learning Research}, vol. 15, no. 1, pp.
  1929--1958, 2014.

\bibitem{tzirakis2018end}
P.~Tzirakis, J.~Zhang, and B.~W. Schuller,
\newblock ``End-to-end speech emotion recognition using deep neural networks,''
\newblock in {\em Proc. of ICASSP}. IEEE, 2018, pp. 5089--5093.

\bibitem{girdhar2017attentional}
R.~Girdhar and D.~Ramanan,
\newblock ``Attentional pooling for action recognition,''
\newblock in {\em Proc. of NeurIPS}, 2017, pp. 33--44.

\bibitem{busso2008iemocap}
C.~Busso, M.~Bulut, C.~Lee, A.~Kazemzadeh, E.~Mower, S.~Kim, J.~N Chang,
  S.~Lee, and S.~S Narayanan,
\newblock ``Iemocap: Interactive emotional dyadic motion capture database,''
\newblock {\em Language resources and evaluation}, vol. 42, no. 4, pp. 335,
  2008.

\bibitem{satt2017efficient}
A.~Satt, S.~Rozenberg, and R.~Hoory,
\newblock ``Efficient emotion recognition from speech using deep learning on
  spectrograms.,''
\newblock in {\em Proc. of INTERSPEECH}, 2017, pp. 1089--1093.

\bibitem{ma2018emotion}
X.~Ma, Z.~Wu, J.~Jia, M.~Xu, H.~Meng, and L.~Cai,
\newblock ``Emotion recognition from variable-length speech segments using deep
  learning on spectrograms,''
\newblock in {\em Proc. of INTERSPEECH}, 2018, pp. 3683--3687.

\bibitem{kingma2015adam}
D.~P. Kingma and J.~L. Ba,
\newblock ``Adam: a method for stochastic optimization,''
\newblock in {\em Proc. of ICLR}, 2015.

\end{thebibliography}

\end{document}